\documentclass[aps,prb,twocolumn]{revtex4}

\usepackage{graphicx,amssymb,amsfonts,amsmath,graphicx}

\newcommand{\intr}{\int\!\!d^2r}
\newcommand{\intq}{\int\!\!\frac{d^2q}{(2\pi)^2}}
\newcommand{\vu}{{\bf u}}
\newcommand{\vq}{{\bf q}}
\newcommand{\ud}{\dot{{\bf u}}}
\newcommand{\ul}{u_L}
\newcommand{\ut}{u_T}
\newcommand{\vl}{v_L}
\newcommand{\vt}{v_T}
\newcommand{\uld}{\dot{u}_L}
\newcommand{\utd}{\dot{u}_T}
\newcommand{\vld}{\dot{v}_L}
\newcommand{\vtd}{\dot{v}_T}
\newcommand{\uldd}{\ddot{u}_L}
\newcommand{\utdd}{\ddot{u}_T}

\newcommand{\uto}{u_{T_0}}
\newcommand{\ulo}{u_{L_0}}
\newcommand{\pto}{p_{T_0}}
\newcommand{\plo}{p_{L_0}}

\newcommand{\wpl}{\omega_+}
\newcommand{\wmi}{\omega_-}
\newcommand{\wl}{\omega_L}
\newcommand{\wt}{\omega_T}
\newcommand{\wc}{\omega_c}
\newcommand{\wsum}{\omega_c^2+\omega_T^2+\omega_L^2}

\newcommand{\Ol}{O_L}
\newcommand{\Ot}{O_T}

\newcommand{\Wpl}{\Omega_+}
\newcommand{\Wmi}{\Omega_-}
\newcommand{\Wl}{\Omega_L}
\newcommand{\Wt}{\Omega_T}
\newcommand{\Wsum}{\omega_c^2+\Omega_T^2+\Omega_L^2}
\newcommand{\epwp}{e^{i\omega_+ t}}
\newcommand{\emwp}{e^{-i\omega_+ t}}
\newcommand{\epwm}{e^{i\omega_- t}}
\newcommand{\emwm}{e^{-i\omega_- t}}

\newcommand{\ua}{{\bf u}_A}

\newcommand{\ub}{{\bf u}_B}

\newcommand{\Ca}{c_A}
\newcommand{\Cb}{c_B}
\newcommand{\Cda}{c_A^\dagger}
\newcommand{\Cdb}{c_B^\dagger}

\newcommand{\suma}{\underset{\alpha}{\sum}}
\newcommand{\sums}{\underset{s}{\sum}}

\newcommand{\ea}{\epsilon_A}
\newcommand{\eb}{\epsilon_B}

\begin{document}
\title{Tunneling between two-dimensional electron systems in a high magnetic
field: role of interlayer interactions}
\author{F. D. Klironomos}
\author{Alan T. Dorsey}
\affiliation{Department of Physics, University of Florida, P.O. Box 118440,
Gainesville, Florida 32611-8440}
\date{\today}

\begin{abstract}
We calculate the tunneling current for a bilayer quantum Hall system
in the interlayer incoherent regime. In order to capture the strong
correlation effects we model the layers as two Wigner crystals
coupled through interlayer Coulomb interactions, treated in the
continuum limit. By generalizing previous work by Johansson and
Kinaret (JK), we are able to study the effect of the low energy
out-of-phase magnetophonon modes on the electron ``shake-up'' which
occurs during a tunneling event. We find the tunneling current peak
value to scale with the magnetic field as found by JK; however, we
find a different scaling of the peak value with the interlayer
separation, which agrees with the measurements by J. P. Eisenstein
{\it et al.}, {\it Phys. Rev. Lett.} {\bf 74}, 1419 (1995).
\end{abstract}

\maketitle

\section{Introduction}
As a consequence of interlayer interactions, bilayer two dimensional
electron systems can exhibit novel behaviors in the high magnetic
field (quantum Hall) regime. The virtue of the bilayer system is
that the relative strength of the intralayer to the interlayer
interactions can be varied; their relative strengths are
characterized by the dimensionless quantity $d/l$, with $d$ the
interlayer separation and $l$ the magnetic length. For example, a
balanced bilayer system ({\it i.e}, with equal densities) at total
filling factor $\nu_T=1/2$ (each layer in a $\nu=1/4$ state) or
$\nu_T=1$ (each layer in a $\nu=1/2$ state) exhibits quantum Hall
plateaus which would have been forbidden for the single layer
system.\cite{suen92,murphy94} A spectacular manifestation of
interlayer interactions for $\nu_T=1$ is the onset of spontaneous
interlayer coherence, which occurs for closely spaced layers (with
$d/l<1.8$). The interlayer coherent regime is characterized by a
prominent zero bias peak in the tunneling
conductance,\cite{spielman00} which disperses with the application
of an in-plane magnetic field.\cite{spielman01} Recent transport
studies provide compelling evidence that the coherent state can be
understood as a bilayer excitonic condensate.\cite{kellogg04,tutuc04}

Here we will focus instead on the tunneling properties in the
\textit{incoherent} regime. Even in the absence of interlayer
coherence there are still important effects of the interlayer
interactions, in particular the Coulomb barrier
peak\cite{spielman00,eisenstein92} which is observed in the
tunneling current-voltage characteristic (I--V). This behavior is a
consequence of both the interactions and the electron dynamics in
high magnetic fields. A semiclassical picture is that the electrons
in the accepting layer accommodate a tunneling electron by forming a
correlation hole. However, the cyclotron motion of the electrons in
a high magnetic field frustrates the formation of this hole (rather
than moving radially outward to form the hole the electrons in the
accepting layer precess, due to the Lorentz force), and the
tunneling can only proceed if the tunneling electrons are externally
provided with sufficient energy to overcome this ``Coulomb
barrier''. The result is a peak in the tunneling current at a bias
equal to the typical Coulomb energy, such that $eV\approx
e/4\pi\epsilon a_0$, where $a_0$ is the average spacing between
electrons. The width and shape of the I--V characteristic will
depend upon the interplay of intralayer versus interlayer
interactions ({\it i.e.} on the parameter $d/l$), and is the subject
of this paper. The semiclassical picture described above is
essentially a dynamical version of a static exciton picture of the
tunneling process, first discussed by Eisenstein {\it et
al.},\cite{eisenstein95}, in which the tunneling electron and the
hole that it leaves behind combine to form an exciton; hints of the
exciton's existence emerge in shifts of the I--V characteristics
with interlayer spacing (the Coulomb barrier peak is ``red-shifted''
to a lower potential by virtue of the interlayer Coulomb energy
$-e^2/\epsilon d$).

To study the tunneling current we adopt a model proposed by
Johansson and Kinaret\cite{johansson94} (JK), in which the two
dimensional electron system in each layer is assumed to have formed
a Wigner crystal (WC). Of course, for the experimentally realized
filling factors the layers are in a liquid phase; the JK model is
only intended to capture the important short distance correlations
among the electrons, and the magnetophonon modes of the WC provide a
reservoir of low energy excitations that can dissipate the energy of
the tunneling electron. In the approach of JK the two layers were
assumed to be noninteracting; we depart from their approach by
incorporating the neglected interlayer interactions. These
interactions \textit{qualitatively} change the collective mode
structure of the bilayer system by introducing a gap in the out of
phase magnetophonon mode and by affecting dramatically the I--V
characteristics of the bilayer system since, as we mentioned
earlier, they ``red-shift" the peak bias values by an amount
proportional to the excitonic energy $e^2/\epsilon d$. The tunneling
electron couples to the out-of-phase magnetophonon mode, and the
electronic ``shake-up'' which accompanies the tunneling event
depends upon the behavior of this mode.

This paper is organized as follows. In Sec. II we review the
continuum theory of the single layer WC. In Sec. III we generalize
this to the bilayer case, including both long range and short range
correlations between the two layers, and find the collective modes
for the bilayer WC. We also determine the coupling between a
tunneling electron and the collective modes of the bilayer WC, which
can be written as a type of spin-boson model.\cite{leggett87}
Section IV is devoted to the calculation of the tunneling current,
which is expressed in terms of a correlation function that is
determined by solving a linear integral equation both numerically
and with an approximate analytical method. The solution allows us to
qualitatively reproduce the tunneling current curves and obtain a
good theoretical estimate for the peak bias values. It also reproduces
the expected peak bias voltage dependence of the I--V curve with
interlayer distance. Many of the details of our calculations
are provided in the appendices.

\section{Single layer theory}

We will start by introducing the single layer model, in which the
2-D electron system is assumed to be in a Wigner crystal state,
modeled as an elastic medium with a momentum cut-off of
$q_0=2\sqrt{\pi n_0}$, where $n_0$ is the single layer density (the
cut-off is determined by imposing conservation of the total number
of states). Of course, for the filling factors relevant to the
tunneling experiments this is not the ground state of the true
system. This simplified approach to the highly correlated
liquid--like state of the electron system takes care of the short
range correlations among electrons and creates an abundance of low
energy collective excitations, which are needed to relax the local
defects created by tunneling events. In order to describe the system
as an elastic medium we need to introduce the two Lam\'{e}
coefficients (the compressional and shear moduli) $\lambda$, $\mu$
which are associated with the energy cost of deformations in the
system.\cite{landau} The effect of the perpendicular magnetic field
is introduced through the vector potential ${\bf A}$. The Lagrangian
that describes the dynamics of the 2-D electron system in the
presence of the perpendicular magnetic field and the intralayer
Coulomb interaction is given by
\begin{widetext}
\begin{equation}
L=n_0\intr \biggl\{\frac{1}{2}m\ud^2-e\ud\cdot{\bf
A}(\vu)-\frac{\lambda}{2n_0} (\partial_i
u_i)^2-\frac{\mu}{4n_0}(\partial_m u_l+\partial_l u_m)^2
+\frac{1}{2}n_0\intr'
 \frac{e^2 [\nabla\cdot\vu({\bf r})][\nabla'\cdot\vu({\bf
r}')] }{4\pi\epsilon|{\bf r}-{\bf r}'|}\biggr\},
\label{r-init-lag}
\end{equation}
\end{widetext}
where $\epsilon$ is the dielectric constant of the host
materials (GaAs in this case) and $\vu$ is the electron lattice
displacement field. Notice that for the intralayer Coulomb interaction
term we have used the linear continuous approximation for charge
density fluctuations $\delta n/n_0 = - \nabla\cdot {\bf u}$ (this is
correct in the absence of vacancies and interstitials). We choose to
work in the symmetric gauge ${\bf A}(\vu)=(-Bu_y/2\;,\;Bu_x/2\;,\;
0)$ where $B$ is the applied magnetic field. In the absence of $B$
the eigenmodes of the elastic medium can be labeled as either
longitudinal or transverse; the magnetic field couples those two
modes together and one has to diagonalize the Hamiltonian of the
system anew. Nevertheless it is useful to decompose the displacement
field $\vu$ into transverse and longitudinal components $u_T$ and $u_L$.
After Fourier--transforming, the above Lagrangian becomes
\begin{align}
L&=n_0\intq\bigg[\frac{1}{2}m\utd^2+\frac{1}{2}m\uld^2+\frac{1}{2}m\wc[\utd\ul
-\uld\ut]\notag\\
&-\frac{1}{2}m\wl^2\ul^2-\frac{1}{2}m\wt^2\ut^2\bigg], \label{q-lag}
\end{align}
where we have defined
\begin{equation}
\wl=\sqrt{c_L^2q^2+\frac{e^2n_0}{2m\epsilon}q}, \quad \wt=c_T|\vq|,
\label{modes}
\end{equation} as the longitudinal and transverse zero
magnetic field eigenfrequencies with
\begin{equation}
c_L=\sqrt{(\lambda+2\mu)/mn_0}, \quad c_T=\sqrt{\mu/mn_0},
\end{equation}
and $\wc=eB/m$ is the cyclotron frequency. If we compare these modes
with the results of Bonsall and Maradudin\cite{bonsall77} for a WC
with pure Coulomb interactions, we see that $c_L$ is a higher order
correction to the magnetoplasmon contribution (the second term on
the right hand side of Eq. \ref{modes}), and can in most cases be
neglected.

We show in Appendix A that the new eigenfrequencies for the
collective modes of the single layer electron system are given by
\begin{align}
\omega_\pm^2&=\frac{1}{2}\bigg[\wsum\bigg] \nonumber \\
&\pm\frac{1}{2}\bigg[(\wsum)^2-4\wt^2\wl^2\bigg]^{1/2}. \label{wpm-eigenfreq}
\end{align}
In the zero magnetic field limit ($\wc=0$) the above modes decouple
into pure longitudinal and transverse modes, while for $q\rightarrow
0$, the magnetoplasmon mode $\omega_+\rightarrow \omega_c$, in
accordance with Kohn's theorem.\cite{kohn61} For $\wc\gg \omega_L,\
\omega_T$, the eigenmode frequencies can be expanded as
\begin{gather}
\wpl=\wc+\frac{\wl^2+\wt^2}{2\wc}+O(\wc^{-3}), \label{wp-wcinf}\\
\wmi=\frac{\wl\wt}{\wc}+O(\wc^{-3}).\label{wm-wcinf}
\end{gather}
The magnetophonon mode $\omega_- \sim q^{3/2}$ at long wavelengths,
in agreement with Bonsall and Maradudin.\cite{bonsall77}

\section{Bilayer theory}

We now consider a bilayer system composed of two parallel single
layers \textit{A} and \textit{B}, separated by a distance $d$, with
the magnetic field normal to the layers. In a continuum model, the
interlayer Coulomb interactions include a long range term which
accounts for interactions between density fluctuations in the two
layers, and a short range term which accounts for the commensuration
energy between the WCs in the two layers. The latter term imposes an
energy penalty for out-of-phase fluctuations of the electron
densities in the two layers, proportional to $(\ua-\ub)^2$. The
Lagrangian for the bilayer system is then
\begin{eqnarray}
L& = &L_A+L_B+n_0\intr\bigg[-\frac{1}{2}\frac{K}{n_0}(\ua-\ub)^2 \nonumber \\
& & -n_0\intr' \frac{e^2 [\nabla\cdot\ua({\bf r})][\nabla'\cdot\ub({\bf r}')]}
{4\pi\epsilon\sqrt{(x-x')^2+(y-y')^2+d^2}}\biggr],
\label{bilayer-lag}
\end{eqnarray}
where $L_A$, $L_B$ are the individual single layer Lagrangians given
by Eq. (\ref{q-lag}), and $K$ is a ``spring constant'' for the
out-of-phase fluctuations. When written in terms of the in-phase
${\bf v}=(\ub+\ua)/2$ and out-of-phase $\vu=\ub-\ua$ field
displacement modes, the Lagrangian decouples into independent
in-phase and out-of-phase terms, so that there are two in-phase and
two out-of-phase eigenmodes; the details are included in Appendix B.
The in-phase modes correspond to a uniform translation of both
layers, and are unimportant for the tunneling processes which are
the topic of this paper. The dispersion of the out-of-phase modes is
similar to Eq. (\ref{wpm-eigenfreq}) and given by
\begin{gather}
\begin{split}
\Omega_\pm^2&=\frac{1}{2}\bigg[\Wsum\bigg]\\
&\pm\frac{1}{2}\bigg[(\Wsum)^2-4\Wl^2\Wt^2\bigg]^{1/2}, \label{Wpm-eigenfreq}
\end{split}
\intertext{where}
\Wt^2=c_T^2q^2+\frac{2K}{mn_0},\\
\Wl^2=c_L^2q^2+\frac{2K}{mn_0}+\frac{e^2n_0}{2m\epsilon}q(1-e^{-qd}).
\end{gather}
Both of these modes are gapped at long wavelengths due to the short
range interlayer coupling $K$.
In the high magnetic field limit we obtain for the
magnetoplasmon ($\Wpl$) and magnetophonon ($\Wmi$)
\begin{gather}
\Wpl=\wc+\frac{\Wl^2+\Wt^2}{2\wc} + O(\omega_c^3),\\
\Wmi=\frac{\Wl\Wt}{\wc} + O(\omega_c^3). \label{magnetophonons}
\end{gather}
The fact that the magnetophonon mode is now gapped, contrary to
the single layer case in which is gapless, is an important distinction between the
single layer and bilayer case for the tunneling characteristics.

\section{Tunneling current}

So far we have developed an elastic theory for the electronic charge
fluctuations in the bilayer system, whose low energy
excitations can provide an energy dissipating mechanism necessary
for the system to relax any charge defect created by tunneling
events. What remains to be done is to couple these collective modes
to the tunneling electrons in the system. Following  JK we assume
that the tunneling events are sufficiently infrequent that we may
treat them as occurring independently, and hence develop a model of
a single tunneling electron which is coupled to the WCs in each
layer. Such a model can incorporate the effects of interlayer
interactions (through the bilayer collective modes), but not
interlayer coherence. The resulting coupling of a tunneling electron
with the magnetophonons (the bilayer charge density fluctuations) is
a standard electron--phonon coupling and is further discussed in
Appendix C. The final form of the Hamiltonian is
\begin{equation}
H=H_0 + H_{\rm bath} +H_T^+ + H_T^-, \label{hamiltonian}
\end{equation}
where
\begin{align}
 H_0&=\bigg[\ea+i\sums M_{sA}\big(a_s^\dagger-a_s\big)\bigg]\Cda\Ca \notag\\
&+\bigg[\eb+i\sums M_{sB}\big(a_s^\dagger-a_s\big)\bigg]\Cdb\Cb, \label{e-ph-coup}\\
&\phantom{====}H_{\rm bath}=\sums\hbar\Omega_s a_s^\dagger a_s,\\
&\phantom{=}H_T^+=T\Cda\Cb, \quad H_T^-=T\Cdb\Ca .
\end{align}
In the above $M_{sA(B)}$ are the electron--magnetophonon couplings and
$\epsilon_{A(B)}$ the corresponding Madelung energies for the two
Wigner crystal lattices, $c$ and $c^\dagger$ are the tunneling electron
annihilation and creation operators, $a_s$ and $a_s^\dagger$ are the
annihilation and creation operators for the collective modes, and $T$ is the
interlayer tunneling matrix element.\cite{bardeen61} The collective
modes that couple to a tunneling electron provide a mechanism to
generate a ``shake-up" that will relax the defect it has produced
and dissipate the extra bias energy the tunneling electron has
acquired.

The tunneling current related to the above Hamiltonian will be
derived from Fermi's golden rule and has the following form\cite{johansson94}
\begin{eqnarray}
I(V)&=&\frac{e}{\hbar^2}\int_{-\infty}^{+\infty}\!\!\!\!\!\!dt
e^{ieVt/\hbar}\langle
\big[H_T^-(t),H_T^+(0)\big]\rangle \nonumber \\
&=&\frac{e}{\hbar^2}\int_{-\infty}^{+\infty}\!\!\!\!\!\!dt\bigg[e^{\frac{ieVt}{\hbar}}I^{\mp}(t)
-e^{-\frac{ieVt}{\hbar}}I^{\pm}(t)\bigg]
\end{eqnarray}
where the correlation function definitions are
\begin{eqnarray}
I^{\mp}(t) & = &\langle H_T^-(t)H_T^+(0)\rangle, \nonumber
\\
I^{\pm}(t)&=&\langle H_T^+(t)H_T^-(0)\rangle,
\label{c+-}
\end{eqnarray}
and the time--dependence is meant in the interaction picture
representation, namely
\begin{equation}
H_T^\pm(t)=e^{\frac{i}{\hbar}H_0t}H_T^\pm
e^{-\frac{i}{\hbar}H_0t}.
\end{equation}
For the calculation of the correlation functions in Eq. (\ref{c+-})
we use the same approach as JK. We assume the tunneling process is
statistically independent from the collective mode propagation and
can thus be averaged independently. For the statistical averaging of
the collective modes we use the linked cluster expansion
method\cite{mahan} which turns out to involve only one link in the
exponential resummation. In Appendix D we show in more detail
how the calculation proceeds. For the correlation function we obtain
\begin{gather}
I^{\mp}(t)=\nu(1-\nu)T^2C(t)\label{I-+t}\\
\intertext{where}
\begin{split}
C(t)&=\exp\bigg\{-\sums\frac{(M_{sB}-M_{sA})^2}{(\hbar\Omega_s)^2}\\
&\times\bigg[\big(N_s+1\big)\big(1-e^{-i\Omega_st}\big)+N_s\big(1-e^{i\Omega_st}
\big)\bigg]\bigg\},
\end{split}\label{c-t}
\end{gather}
and $N_s$ is the boson thermal occupation number for the
collective modes of the system. To get the form of Eq.
(\ref{c+-}) it suffices to interchange $A$ and $B$ in Eq.
(\ref{c-t}). Notice that the in-phase modes drop out from
the above correlation function due to their matrix element
property $M_{sA}=M_{sB}$, this is not the case for the
out-of-phase modes which, as we show in the Appendix, obey
$M_{sA}=-M_{sB}$.

The experimental temperature range in the tunneling current results
is of the order of 0.1 K$\sim$10$^{-5}$ eV while the bias voltage is
in the range of mV, so a zero temperature calculation is
appropriate; this simplifies things considerably since the bosonic
occupation numbers $N_s=0$. In addition, due to the high magnetic
field the magnetoplasmon modes will have a large gap and will not
contribute to the electron coupling. Gathering all this together
and switching to dimensionless units for the momentum integration
($x=q/q_0$) we find the following result for the time dependent
correlation function
\begin{equation}
C(t)=\exp\bigg[\int_0^1dxf(x)\big(e^{-i\omega(x)t}-1\big)\bigg].
\label{C_t_T0}
\end{equation}
In Appendix D we provide the analytic expression for the
weight--function $f(x)$, while $\omega(x)$ is given by Eq.
(\ref{magnetophonons}) or specifically by Eq. (\ref{omega_x}). By
differentiating Eq. (\ref{C_t_T0}) and taking the Fourier transform
we obtain the following equation for the Fourier transform of the
correlation function
\begin{equation}
\omega C(\omega)=\int_0^1dxf(x)\omega(x) C(\omega-\omega(x)).
\label{i-eq-c}
\end{equation}
As we show in Appendix D this correlation function is zero for
$\omega\le 0$.

The integral equation in Eq. (\ref{i-eq-c}) is difficult to solve
analytically. However, we can find the asymptotic behavior of
$C(\omega)$ by expanding in $\omega(x)$. To lowest order we
obtain a first order differential equation with the solution
\begin{equation}
C(\omega)\sim\exp\bigg[-\frac{(\omega-c_1)^2}{2c_2}\bigg],\label{asymptotic}
\end{equation}
where we provide analytic expressions for $c_1$ and $c_2$ in
Appendix D. This expansion gives us an analytic prediction for the
peak bias voltage, namely $c_1$, and the general behavior of the
correlation function at large bias voltages. We can investigate the
behavior of $c_1$ in the two limiting cases when the interlayer
Coulomb interaction is much weaker than the intralayer one and vice
versa. We find that for the $d\gg a_0$ limit, $c_1$ scales as
$1/a_0$, and the width of the Coulomb barrier peak $\sqrt{c_2}$
scales as $\sqrt{c_2}\sim 1/a_0^2\sqrt{B}$. This is the same
phenomenological behavior JK extract from their calculations. In the
opposite limit $d\ll a_0$, we find that $c_1\sim d^2/a_0^3$ and
$\sqrt{c_2}\sim d/a_0^3\sqrt{B}$. This limit is absent in the JK
model. We should notice here that in this regime the actual behavior
of the system will be significantly modified by coherence effects
but those effects will be pronounced at zero-bias. At finite bias
values we see from experimental results\cite{spielman00} that the
Coulomb barrier peak survives, but it is ``red-shifted"
significantly which, as we see, our model is able to reproduce as a
limiting behavior.

\begin{figure}[t]
\centering
\includegraphics[totalheight=8cm,width=8cm,
viewport=0 0 680 500,clip]{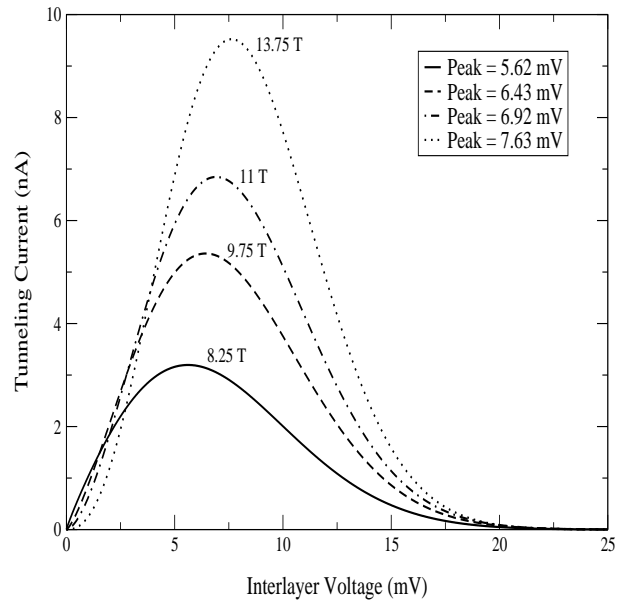} \caption{Tunneling
current curves for different magnetic field values using the moment
expansion solution of Eq. (\ref{i-eq-c}). The legend shows the peak
bias values calculated by Eq. (\ref{peak}).} \label{m.results}
\end{figure}

The importance of the asymptotic result lies on the insight it
provides on the tunneling current behavior. We use this information
to build an \emph{Ansatz} for the correlation function. We are after
a qualitative analytic expression that will capture the basic
physics and reproduce the Gaussian asymptotic behavior. We choose
the following \emph{Ansatz}
\begin{equation}
C(\omega)=N\omega^re^{-\lambda\omega^2}.\label{ansatz}
\end{equation}
As we show in Appendix E, this choice reproduces very accurately
the numerical solution of the equation which is our strongest
justification for using it. For the parameters involved in this
choice of $C(\omega)$ we develop a self consistent method of
evaluation. This method is based on the moment expansion of
$C(\omega)$. If we multiply--differentiate Eq. (\ref{C_t_T0}) we end
up to the following moment equations
\begin{gather}
\int_0^\infty d\omega C(\omega)=2\pi, \label{i0}\\
\int_0^\infty d\omega\omega C(\omega)=2\pi c_1, \label{i1}\\
\int_0^\infty d\omega\omega^2C(\omega)=2\pi\big(c_2+c_1^2\big).\label{i2}
\end{gather}
These are the three equations that our Ansatz should obey
self--consistently. We show in Appendix E in more detail how we
extract the values for $N$, $r$ and $\lambda$ from them. Our
theoretical value for the peak bias voltage can be easily calculated
now from Eq. (\ref{ansatz}). The result is
\begin{equation}
V_0=\frac{1000\hbar}{e}\sqrt\frac{r}{2\lambda}\label{peak}
\end{equation}
where we have converted it to mV. In Fig. (\ref{m.results}) we present the results
the moment expansion method produces. The theoretical parameters are directly taken
from the experiment.\cite{eisenstein92} The bilayer sample area is
$S$=0.0625mm$^2$ and the single layer electron density is $n_0$=1.6 $\times$
10$^{11}$ cm$^{-2}$. The perpendicular magnetic field varies from 8 T to 13.75 T and
the Wigner crystal lattice parameter has the value $a_0\simeq$ 270 \AA{}
which corresponds to the stable hexagonal lattice configuration
($n_0=2/\sqrt{3}a_0^2$).\cite{bonsall77}
The double well separation distance is $d$=175 {\AA} and the dielectric constant of
GaAs is $\epsilon$=12.9$\epsilon_0\simeq$1.14 $\times$ 10$^{-11}$ F/m. The
transverse sound velocity for the electron gas is given by\cite{bonsall77}
\begin{equation}
c_T\simeq\sqrt{0.0363\frac{e^2}{\sqrt{3}\epsilon m a_0}}\simeq 53552\textrm{ m/s.}
\label{ct}
\end{equation}
For the electron mass we use the electron effective mass value in the GaAs background 
$m$=0.067$m_e$. For the $K$ parameter we notice that by construction $K/n_0$ is a measure 
of the energy density per electron due to the short range part of the interlayer Coulomb
interactions. So we assume it will scale accordingly as
\begin{equation}
\frac{K}{n_0}=\kappa\frac{e^2/4\pi\epsilon d}{\pi l^2},\;\kappa>0.
\label{kappa}
\end{equation}
The value of $\kappa$ is a measure of the magnetophonon gap in this system.
C\^ot\'e and collaborators have performed time--dependent Hartree
Fock calculations investigating the magnetophonon dispersion
relation and crystalline phases for the bilayer system.\cite{cote}
They were able to provide us with a $\kappa$=0.0085 value. With this last piece
of the puzzle in place we are able to produce our analytic results for the tunneling
current shown in Fig. (\ref{m.results}).
As we see, the model captures the qualitative behavior of the current and it 
accurately reproduces the experimental results for the peak bias voltages associated
with different applied magnetic field values. Notice though, that the pronounced 
experimental low bias suppression\cite{eisenstein92} is lifted in our case although
present in the JK model\cite{johansson94} and in spectral weight calculations\cite{He93}
associated with this bilayer system. We attribute this behavior to the very small 
value of $\kappa$ and mostly to the role of interlayer interactions. In particular,
as it can be seen in  Fig. (\ref{m.results}), the Coulomb blockade tends to be 
restored with increasing magnetic field because the role of interlayer interactions
is diminished in that limit. We observe the same behavior as well when the interlayer
distance $d$ is increased, which signifies the same physical limit, where essentially
the JK results are recovered. We believe that in order to capture all quantitative 
characteristics of the tunneling current for the whole range of bias voltages one 
has to go beyond the continuum approximation employed here.

Another way of calculating the correlation function is by numerically integrating 
Eq. (\ref{i-eq-c}). In Appendix E we report analytically on this approach. The results
we get are shown in Fig. (\ref{numerical.results}) and are very similar to the moment
expansion results; the same applies for the peak bias values.

\begin{figure}[t]
\centering
\includegraphics[totalheight=7cm,width=8cm,
viewport=0 0 820 705,clip]{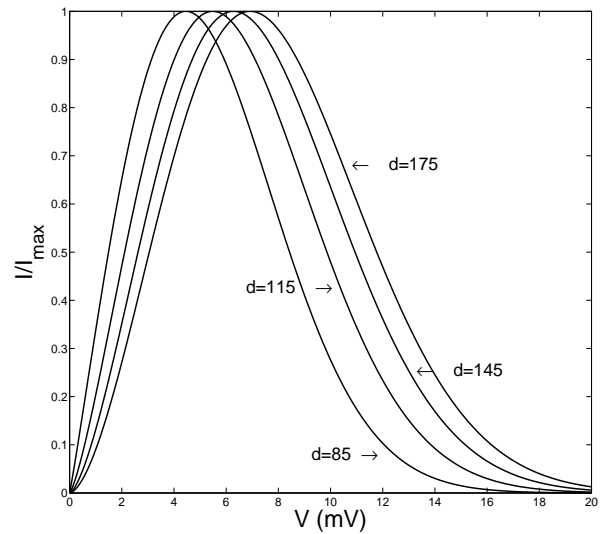}\caption{Normalized tunneling current
curves for different interlayer separation distances $d$ measured in \AA.} \label{IV-d}
\end{figure}
At this point we are in position to analytically investigate how our
solution behaves for different interlayer separation distances $d$.
Notice that the dependence in $d$ is introduced in two places in our
model, the one is the tunneling matrix elements $T$ (which depends
exponentially on $d$) and the other is through the long-range part
of the interlayer Coulomb interaction. Since we are interested in
the latter only, we will normalize the I--V curves for different $d$
values to the peak current value (this has the effect of removing the
tunneling matrix element contribution). According to experimental
results\cite{eisenstein95} we expect to see a ``red-shift" in the
I--V curves associated with the bounding interlayer exciton energy
$-e^2/\epsilon d$. Our results are shown in Fig. (\ref{IV-d}) and as
we see our model captures this kind of important interlayer
correlation physics. According to our model, we can conclude that
the effect of the interlayer Coulomb interaction in the real system
is two-fold. First, the short-range correlations derived from the
interlayer interaction introduce a gap in the low frequency
excitations, which is well pronounced in the experimental results,
\cite{eisenstein92} and second, the long-range effects of the
interlayer interactions ``soften" the response of the bilayer system
to tunneling, through the exciton creation associated with tunneling
events.

\section{Summary}

We have introduced a model for tunneling in a bilayer system in high
magnetic fields, which explicitly includes the interlayer
interactions. The bilayer system of electrons is modeled as two
correlated Wigner crystals in the continuum limit. The low energy
collective modes of the crystals are coupled to the tunneling
electrons in the system, providing an energy dissipation mechanism
to relax the charge defect produced by the tunneling event.
This model allows for an approximate analytic solution to the
tunneling current problem that captures the basic experimental
behavior of the tunneling current curve and fairly accurately
reproduces the peak bias values for different applied magnetic field
cases along with the expected I--V dependence with interlayer distance $d$.
In the future we would like to find a way to include
coherence into the model and reproduce the zero bias peak
anomaly\cite{spielman00} the tunneling current exhibits.

\section{Acknowledgments}

We would like to thank S. M. Girvin for valuable discussions, and R.
C\^ot\'e and his collaborators for providing us with numerical
values for the magnetophonon gap, and J. Eisenstein for helpful
comments on an earlier version of the manuscript.
This work was supported by NSF DMR-9978547.

\appendix

\section{Single layer eigenmodes}

We have expressed the single layer dynamics by writing down the Lagrangian of
Eq. (\ref{q-lag}). The canonical quantization process is a textbook exercise where
one has to solve the Lagrange equation of motion for the fields, namely
\begin{align}
\utdd+\wc\uld+\wt^2\ut=0, \\
\uldd-\wc\utd+\wl^2\ul=0.
\end{align}
Notice the effect that the magnetic field has in coupling the otherwise orthogonal
longitudinal and transverse zero field eigenmodes. The above system of equations is easy to
prove that has eigenvalues given by Eq. (\ref{wpm-eigenfreq}).
The general solution for the eigenmodes is of the form
\begin{gather}
\ut=A_T\epwp \!+\! A'_T\emwp \!+\! B_T\epwm \!+\! B'_T\emwm, \\
\begin{split}
\ul&=i\frac{\wc\wpl}{\wpl^2-\wl^2}[-A_T\epwp+A'_T\emwp]\\
&+i\frac{\wc\wmi}{\wmi^2-\wl^2}[-B_T\epwm+B'_T\emwm],
\end{split}
\end{gather}
where $A_T$, $A_T'$ and $B_T$, $B_T'$ are the coefficients
associated with the creation and annihilation of the two modes
respectively. These are directly related to the creation and
annihilation operators by the normalization condition. The way we
reach to the latter is by expressing all the higher derivative terms
in terms of the canonical displacements and the canonical momenta at
time equal to zero which obey the commutation relations
$[u_{i_0}(\vq),p_{j_0}(\vq')]=i\hbar\delta_{ij}(2\pi)^2\delta^2(\vq-\vq')$.
After a fair amount of algebra we have convinced ourselves that the operators obey the
correct eigenmode commutation relation. By applying the
normalization condition we can reach the final form of the eigenmode
operators given by (implicit $\vq$ dependence will be suppressed unless necessary)
\begin{eqnarray}
a_1(\vq)&=&\!\!\tilde{a}_1\!
\bigg[\frac{\wc\wpl}{n_0m}\plo \!-\! i\wc\frac{\wpl^2+\wl^2}{2}\ulo \!+\! i\frac{\wpl^2-\wl^2}
{n_0m}\pto \nonumber \\
&+&\!\!\wpl\!\bigg(\wpl^2-\wl^2-\frac{\wc^2}{2}\bigg)\uto\bigg],\label{a1}
\end{eqnarray}
\begin{eqnarray}
a_2(\vq)&=&\!\!\tilde{a}_2\!
\bigg[\!\!-\!\frac{\wc\wmi}{n_0m}\plo \!\!+\! i\wc\frac{\wmi^2+\wl^2}{2}\ulo \!\!+\! i\frac{\wl^2-\wmi^2}
{n_0m}\pto \nonumber \\
&+&\!\!\wmi\!\bigg(\wl^2-\wmi^2+\frac{\wc^2}{2}\bigg)\uto\bigg],\label{a2}
\end{eqnarray}
where they obey $[a_i(\vq),a_j^\dagger(\vq')]=(2\pi)^2\delta^2(\vq-\vq')\delta_{ij}$ and
\begin{gather}
\tilde{a}_1=\sqrt{\frac{mn_0}{2\hbar\wpl(\wpl^2-\wl^2)(\wpl^2-\wmi^2)}},\\
\tilde{a}_2=\sqrt{\frac{mn_0}{2\hbar\wmi(\wl^2-\wmi^2)(\wpl^2-\wmi^2)}}.
\end{gather}
We can finally express the displacement fields in terms of the eigenmodes in
Schr\"odinger's representation as
\begin{eqnarray}
\ut(\vq)&=&\sqrt{\frac{\hbar}{2mn_0(\wpl^2-\wmi^2)}}\bigg[\sqrt{\frac{\wpl^2-\wl^2}{\wpl}}
\big(a_1^\dagger+a_1\big) \nonumber \\
&+&\sqrt{\frac{\wl^2-\wmi^2}{\wmi}}\big(a_2^\dagger+a_2\big)\bigg],\label{ut}
\end{eqnarray}
\begin{eqnarray}
\ul(\vq)&=&i\sqrt{\frac{\hbar\wc^2}{2mn_0(\wpl^2-\wmi^2)}}\bigg[\sqrt{\frac{\wpl}{\wpl^2-\wl^2}}
\big(a_1-a_1^\dagger\big) \nonumber \\
&-&\sqrt{\frac{\wmi}{\wl^2-\wmi^2}}\big(a_2-a_2^\dagger\big)\bigg],\label{ul}
\end{eqnarray}

\section{Bilayer eigenmodes}

Starting with Eq. (\ref{bilayer-lag}) and introducing the in-phase
${\bf v}=(\ub+\ua)/2$ and out-of-phase $\vu=\ub-\ua$ modes, we find
that the Lagrangian can be written as $L=L_{\rm in} + L_{\rm{out}}$,
with
\begin{eqnarray}
L_{\rm in}&=& n_0\intq \frac{1}{2}(2m)\bigg[\vld^2+ \vtd^2 +\wc(\vtd\vl-\vld\vt) \nonumber \\
&-&\wt^2\vt^2-\Ol^2\vl^2\bigg],
\end{eqnarray}
\begin{eqnarray}
L_{\rm out}&=&n_0 \intq \frac{1}{2}\frac{m}{2}\bigg[\uld^2+\utd^2+\wc(\utd\ul-\uld\ut) \nonumber \\
&-&\Wt^2\ut^2 -\Wl^2\ul^2\bigg],
\end{eqnarray}
where
\begin{gather}
\Wt^2=c_T^2q^2+\frac{2K}{mn_0},\\
\Wl^2=c_L^2q^2+\frac{2K}{mn_0}+\frac{e^2n_0}{2m\epsilon}q(1-e^{-qd}),\\
\Ol^2=c_L^2q^2+\frac{e^2n_0}{2m\epsilon}q(1+e^{-qd}),\\
\Ot=\wt.
\end{gather}
We see that the four modes have decoupled into two in-phase (with
total mass $2m$) and two out-of-phase (with reduced mass $m/2$)
modes that are governed by effective single layer Lagrangian
dynamics. Notice that the in-phase transverse mode remained unchanged.
In order to obtain the analytic results for the
out-of-phase modes displacement field operators we have to redefine
the parameters in Eqs. (\ref{a1}-\ref{ul}) as $m\rightarrow
m/2$, $\wt^2\rightarrow\Wt^2$, $\wl^2\rightarrow\Wl^2$
while for the in-phase mode we have $m\rightarrow 2m$, $\wl^2\rightarrow\Ol^2$.
The new eigenvalues $\Omega_\pm$ will be given by Eq. (\ref{Wpm-eigenfreq}) and a
corresponding one from Eq. (\ref{wpm-eigenfreq}) for $O_\pm$ when we carry out the above
frequency changes. Notice that the mass in the cyclotron frequency formula does not change
in either case.
We will use the operator form of these eigenmodes to calculate below their coupling
matrix elements with the electrons in the system.

\section{Collective mode coupling matrix}

The coupling between a point charge with charge density $n_e({\bf
r})$ and a charge fluctuation $\delta n({\bf r})$ in either of the
two layers is given by
\begin{equation}
\begin{split}
H_{\rm cpl}&=\frac{e^2}{4\pi\epsilon}\intr\int\!\!d^2r_A\frac{n_e({\bf r})
\delta n_A({\bf r}_A)}{|{\bf r}-{\bf r}_A|}\\
&+\frac{e^2}{4\pi\epsilon}\intr\int\!\!d^2r_B\frac{n_e({\bf r})\delta n_B({\bf r}_B)}
{|{\bf r}-{\bf r}_B-{\bf d}|}.
\end{split}
\end{equation}
In the continuum approximation to lowest order we have $\delta
n_A=-n_0\nabla \cdot\ua$, and we can place the electron at the
origin so that  $n_e({\bf r})=\delta^{(2)}({\bf r})$. After
introducing the in-phase and out-of-phase displacement fields and
Fourier transforming, the coupling becomes
\begin{equation}
H_{\rm cpl} = (c_A^\dagger c_A - c_B^\dagger c_B)\intq
\frac{e^2n_0}{4\epsilon q} (1-e^{-qd}) i \vq\cdot\mathbf{u},
\end{equation}
where we have dropped a constant term (which depends upon
$\mathrm{v}$). Next, we use the operator definition for the
out-of-phase displacement field found previously
\begin{gather}
\ul=-if_1a_1^\dagger+if_1a_1-if_2a_2^\dagger+if_2a_2,
\intertext{where}
f_1=\wc\sqrt{\frac{\hbar}{n_0m}\frac{\Wpl}{(\Wpl^2-\Wl^2)(\Wpl^2-\Wmi^2)}},\\
f_2=-\wc\sqrt{\frac{\hbar}{n_0m}\frac{\Wmi}{(\Wl^2-\Wmi^2)(\Wpl^2-\Wmi^2)}},
\end{gather}
and the coupling term assumes the form of Eq. (\ref{e-ph-coup}), where
the matrix elements are given by
\begin{equation}
M_{sA}=-M_{sB}=-\frac{e^2 n_0}{4\epsilon}\big(1-e^{-dq}\big)f_s,
\end{equation}
and $s=1,2$ refers to the two out-of-phase bilayer eigenmodes but any
summation on it is meant to include an integration in $q$ as well.

\section{Correlation function}

To calculate the correlation function given by Eq. (\ref{c+-}) we
use the linked cluster expansion method.\cite{mahan} As it turns out
there is only one link involved in the exponential resummation of
the series. As pointed out by JK, if we assume that the tunneling
events are faster than the relaxation time for WCs, then the thermal
averaging of the magnetomodes and the tunneling electrons is
statistically independent. The correlation function becomes
\begin{align}
I^{\mp}(t)&=T^2\nu(1-\nu)\langle T_t\exp\bigg[-\frac{i}{\hbar}\int_0^tdt'
e^{\frac{iH_it'}{\hbar}}\nonumber \\
&\times (H_f-H_i)e^{-\frac{iH_it'}{\hbar}}\bigg]\rangle
\end{align}
where
\begin{gather}
H_i=\!\ea \!-\! \Delta_A \!+\! \sums\hbar\Omega_s\big(a^\dagger_s \!-\! \frac{iM_{sA}}{\hbar
\Omega_s}\big)\big(a_s \!+\! \frac{iM_{sA}}{\hbar\Omega_s}\big),\\
H_f=\!\eb \!-\! \Delta_B \!+\! \sums\hbar\Omega_s\big(a^\dagger_s \!-\! \frac{iM_{sB}}{\hbar
\Omega_s}\big)\big(a_s \!+\! \frac{iM_{sB}}{\hbar\Omega_s}\big),
\end{gather}
are the Hamiltonians of the system before and after a tunneling
event respectively. Using the linked cluster expansion method we
obtain the result of Eq. (\ref{I-+t}), aside from an overall phase
factor $(\eb-\Delta_B-\ea+\Delta_A)t/\hbar$ which we assume to be
zero. As JK point out, this amounts to a complete Wigner crystal
relaxation after each tunneling event, where the polaron shifts
$\Delta_A-\Delta_B$ defined by
\begin{equation}
\Delta_{A(B)}=\suma\frac{|M_{\alpha A(B)}|^2}{\hbar\omega_\alpha},
\end{equation}
compensate the Madelung energies $\eb-\ea$ of the interstitial
positions. Another point of interest is that as we see from Eq.
(\ref{c-t}) the matrix elements of the electron coupling with the
collective modes of the system enter in the correlation function
formula as a difference, so that the tunneling electron only couples
to the out-of-phase modes. When an electron tunnels from one layer
to the other the remaining electrons in that layer have to fill in
the hole; at the same time the electrons in the receiving layer need
to ``open up" and create a hole for the incoming electron, so we see
that the relative motion of the two charge densities should be
out-of-phase during a tunneling event.

Since the magnetoplasmons have a large gap in high fields we only
need to consider the magnetophonon matrix element, which in the high
magnetic field case will be given by
\begin{align}
f(x)&=\frac{c\wc^5}{c_T^8q_0^6}\frac{x(1-e^{-\gamma x})^2}{\delta+x^2+2\alpha
+\beta x(1-e^{-\gamma x})}\frac{1}{\sqrt{x^2+\alpha}} \nonumber\\
&\times \frac{1}{[\alpha+\beta x(1-e^{-\gamma x})]^{3/2}}\frac{1}{\delta-\alpha-x^2}.
\label{f_w}
\end{align}
For the magnetophonon frequency we will have
\begin{equation}
\omega(x)=\frac{c_T^2q_0^2}{\wc}\sqrt{\alpha+\beta x(1-e^{-\gamma
x})} \sqrt{x^2+\alpha}.\label{omega_x}
\end{equation}
For convenience we have introduced dimensionless units for the
momenta ($x=q/q_0$), and we have set $c_L=0$. The parameters in the
equations above are
\begin{gather}
c=\frac{n_0}{8\pi\hbar
m}\bigg(\frac{e^2}{\epsilon}\bigg)^2,\label{parm_1}\\
\alpha=\frac{2K}{mn_0c_T^2q_0^2}=\frac{\kappa}{2\pi^2mc_T^2}\frac{1}{(lq_0)^2}
\frac{e^2}{\epsilon d},\\
\beta=\frac{e^2n_0}{2\epsilon m c_T^2q_0}, \\
\gamma=dq_0,\\
\delta=\bigg(\frac{\wc}{c_Tq_0}\bigg)^2.
\label{parm_last}
\end{gather}
The parameter $\alpha$ is dimensionless and gives a measure of the
magnetophonon gap.  We have used Eq. (\ref{kappa}) to produce the
final equation that defines it. The parameter $\gamma$ is
dimensionless and measures the relative interaction strength between
the interlayer and intralayer Coulomb interactions. The parameter
$\beta$ is dimensionless as well and does not depend on the Wigner
crystal lattice parameter $a_0$ as it turns out if the dependence of
$c_T$ given by Eq. (\ref{ct}) is taken into account. In order to
derive the result for the asymptotic behavior of the correlation
function we have to Taylor--expand $C(\omega)$ involved in the
integral of Eq. (\ref{i-eq-c}) in powers of $\omega(x)$. It is
straightforward to show that to lowest order the first order
differential equation we obtain has the Gaussian solution given by
Eq. (\ref{asymptotic}). The coefficients involved in the expansion
are given by
\begin{align}
c_1&=\int_0^1dxf(x)\omega(x)\nonumber\\
&=c\frac{\wc^4}{c_T^6q_0^4}\int_0^{1}dx \frac{x(1-e^{-\gamma x})^2}{\delta+x^2
+2\alpha+\beta x(1-e^{-\gamma x})} \nonumber\\
&\times \frac{1}{\alpha+\beta x(1-e^{-\gamma x})}\frac{1}{\delta-\alpha-x^2}, \label{c_1}
\end{align}
\begin{align}
c_2&=\int_0^1dxf(x)\omega^2(x)\nonumber\\
&=c\frac{\wc^3}{c_T^4q_0^2}\int_0^{1}dx \frac{x(1-e^{-\gamma x})^2}{\delta+x^2
+2\alpha+\beta x(1-e^{-\gamma x})}\nonumber\\
&\times \sqrt{\frac{x^2+\alpha}{\alpha+\beta x(1-e^{-\gamma x})}}\frac{1}
{\delta-\alpha-x^2}. \label{c_2}
\end{align}
As we mentioned in the text we can expand to one order higher and analytically solve
the differential equation as well, but the correction will not be very useful
since these results are valid only in the asymptotic region that a lowest order
calculation suffices to catch the essential characteristics. In the $d\gg a_0$ case
we can ignore the exponentials in the integrand of Eqs. (\ref{c_1}, \ref{c_2}) and
as it turns out a $c_1\sim 1/a_0$ and $\sqrt{c_2}\sim1/a_0^2B$ scaling behavior emerges.
In the opposite limit $d\ll a_0$, if we expand the exponentials in the integrand of
Eqs. (\ref{c_1}, \ref{c_2}) we find that $c_1\sim d^2/a_0^3$ and $\sqrt{c_2}\sim d/a_0^3B$.

At this point we would like to justify the assumption that the
correlation function is zero for zero or negative bias voltage
($\omega\leq 0$). By expanding the exponential in Eq. (\ref{i-eq-c})
and interchanging summation with integration we can perform the last
integral that gives a delta function in the frequencies. Since the
magnetophonons are always positive definite (gapped) we see that the
correlation function
\begin{align}
C(\omega)&=\!\!2\pi e^{-\int_0^1\!dxf(x)}\sum_{n=0}^{\infty}\frac{1}{n!}\int_0^1\!\!\!dx_1f(x_1)
\cdots\!\int_0^1\!\!\!dx_nf(x_n) \nonumber \\
&\times\!\!\delta\big(\omega-\omega(x_1)-\cdots-\omega(x_n)\big)
\end{align}
has to be zero for any $\omega$ less than the lowest possible magnetophonon
frequency value.

With this last piece of information about the correlation function
we can proceed now into developing our method of moment expansion. As we have already
justified we use the Ansatz of Eq. (\ref{ansatz}) and the three different moments
of the correlation function given by Eqs. (\ref{i0}--\ref{i2}).
At this point it is convenient to make a change of variable
$\omega=eV/1000\hbar$ to convert the argument of $C(\omega)$ into mV
and define the corresponding parameter in the exponent as
$\Lambda=\lambda\big(\frac{e}{1000\hbar}\big)^2$. By performing the
integrals the three moment equations become
\begin{gather}
\frac{1}{2}N\bigg(\frac{e}{1000\hbar}\bigg)^{r+1}\!\!\!\Lambda^{-\frac{r+1}{2}}
\Gamma\bigg(\frac{r+1}{2}\bigg)=2\pi, \label{sol0}\\
\frac{1}{2}N\bigg(\frac{e}{1000\hbar}\bigg)^{r+2}\!\!\!\Lambda^{-\frac{r+2}{2}}
\Gamma\bigg(\frac{r+2}{2}\bigg)=2\pi c_1, \label{sol1}\\
\frac{1}{2}N\bigg(\frac{e}{1000\hbar}\bigg)^{r+3}\!\!\!\Lambda^{-\frac{r+3}{2}}
\Gamma\bigg(\frac{r+3}{2}\bigg)=2\pi\big(c_2+c_1^2\big).\label{sol2}
\end{gather}
We can divide Eq. (\ref{sol1}) and Eq. (\ref{sol2}) by Eq. (\ref{sol0}) and then equating
the two we reach the self consistent equation for $r$, namely
\begin{equation}
\frac{\Gamma^2\big(\frac{r+2}{2}\big)}{\Gamma\big(\frac{r+3}{2}\big)}=
\frac{\Gamma\big(\frac{r+1}{2}\big)}{1+c_2/c_1^2}.
\end{equation}
We numerically solve the above equation and obtain a value for $r$.  The usual
range of $r$ for the magnetic field values considered is $1/2<r<2$. This can be
regarded as an estimate for the low-bias current power-low behavior.
Having $r$ at hand we can go back and evaluate the rest of the parameters.
The final form of the correlation function becomes
\begin{equation}
C(V)=N\bigg(\frac{e}{1000\hbar}\bigg)^rV^re^{-\Lambda V^2}
\end{equation}
where $V$ measures in mV. To obtain the peak bias value result of
Eq. (\ref{peak}) we find the root of the first derivative of the
above equation.

\section{Numerical integration}

\begin{figure}[b]
\centering
\includegraphics[totalheight=8cm,width=8cm,
viewport=0 0 680 500,clip]{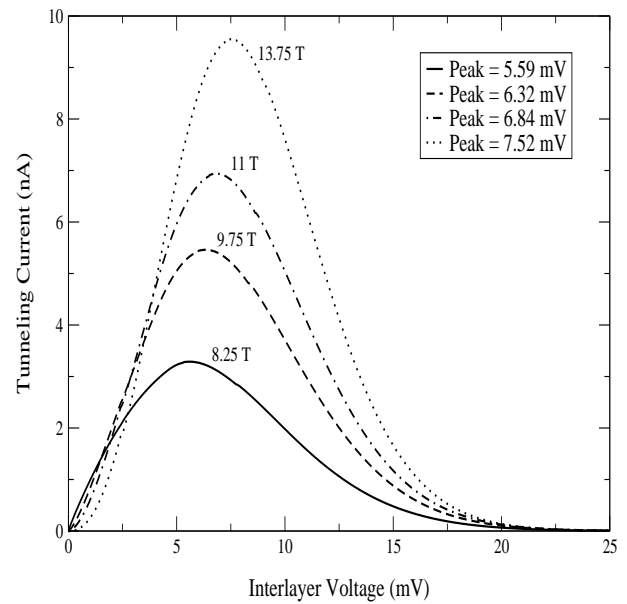} \caption{Tunneling current
curves for different magnetic field values produced by numerically
integrating Eq. (\ref{i-eq-c}). The legend shows the peak bias
values obtained with this approach. They are in close agreement with
the analytic results.} \label{numerical.results}
\end{figure}

We have numerically integrated the integral equation for the
correlation function in two different ways---first by a direct
integration of the integral equation and then by introducing the
density of states (similar to the JK method). As both methods give
the same results we will only report on the latter. The equation to
be integrated is
\begin{equation}
zC(z)=\int_0^1\!\!\!\!dxf(x)\frac{\omega(x)}{\gamma_0}C\big(z-\frac{\omega(x)}{\gamma_0}
\big)\theta\big(z-\frac{\omega(x)}{\gamma_0}\big)
\label{integ_eq}
\end{equation}
where we have introduced the parameter
$\gamma_0=\frac{e}{1000\hbar}$ to convert the frequency argument of
the correlation function into mV. Before we proceed we should notice
that the magnetophonon frequency is bounded in a region
$\alpha_1\leq\frac{\omega(x)}{\gamma_0}\leq\alpha_2$ which means that
the density of states
is non zero only in that range. The values of
$\alpha_1$ and $\alpha_2$ are given by substituting $x=0$ and $x=1$
into Eq. (\ref{omega_x}) respectively. The upper bound $\alpha_2$
appears due to the momentum cutoff we have introduced. The final
form for the correlation function integral equation will be
\begin{equation}
C(z)=\begin{cases}
\frac{1}{z}\int_{\alpha_1}^zdyg(y)C(z-y),&\alpha_1\leq z\leq\alpha_2, \\
&\\
\frac{1}{z}\int_{\alpha_1}^{\alpha_2}dyg(y)C(z-y),& z\geq\alpha_2,
\end{cases}
\end{equation}
where the definition for the density of states is
\begin{equation}
g(y)=y\bigg(\frac{f(x)}{\frac{1}{\gamma_0}|\frac{d\omega(x)}{dx}|}\bigg)_{x(y)},
\end{equation}
and $x(y)$ is the root of the equation $\omega(x)=\gamma_0y$. In
this approach one has to ``jump--start" the algorithm with an
assumption for the low bias points. We can show that for small
frequencies (but greater than the lower magnetophonon bound)
the correlation function will have a $z-\alpha_1$ behavior.
In Fig. (\ref{numerical.results}) we show our results for the
tunneling current using this method. As we see the  qualitative
behavior of the tunneling current is very similar to our analytic
result, and the peak bias values are similar as well.


\begin{thebibliography}{99}


\bibitem{suen92}
Y. W. Suen, L. W. Engel, M. B. Santos, M. Shayegan, and D. C. Tsui, Phys. Rev. Lett.
{\bf 68}, 1379 (1992); J. P. Eisenstein, G. S. Boebinger, L. N. Pfeiffer, K. W. West,
and Song He, Phys. Rev. Lett. {\bf 68}, 1383 (1992).
\bibitem{murphy94} S. Q. Murphy, J. P. Eisenstein, G. S. Boebinger, L. N. Pfeiffer,
and K. W. West,Phys. Rev. Lett. {\bf 72}, 728 (1994).
\bibitem{spielman00}
I. B. Spielman, J. P. Eisenstein, L. N. Pfeiffer, and K. W. West, Phys. Rev. Lett.
{\bf 84}, 5808 (2000).
\bibitem{spielman01} I. B. Spielman, J. P. Eisenstein, L. N. Pfeiffer, and K. W. West,
Phys. Rev. Lett. {\bf 87}, 036803 (2001).
\bibitem{kellogg04} M. Kellogg, J. P. Eisenstein, L. N. Pfeiffer, and K. W. West
Phys. Rev. Lett. {\bf 93}, 036801 (2004).
\bibitem{tutuc04} E. Tutuc, M. Shayegan, and D. A. Huse
Phys. Rev. Lett. {\bf 93}, 036802 (2004).
\bibitem{eisenstein92}
J. P. Eisenstein, L. N. Pfeiffer, and K. W. West Phys. Rev. Lett. {\bf 69}, 3804
(1992).
\bibitem{eisenstein95}
J. P. Eisenstein, L. N. Pfeiffer, and K. W. West, Phys. Rev. Lett.
{\bf 74}, 1419 (1995)
\bibitem{johansson94}
P. Johansson and J. M. Kinaret, Phys. Rev. B {\bf 50}, 4671 (1994).
\bibitem{leggett87} A. J. Leggett, S. Chakravarty, A. T. Dorsey, A. Garg, M. P. A. Fisher,
and W. Zwerger, Rev. Mod. Phys. {\bf 59}, 1 (1987).
\bibitem{landau}
See for example $\S$4 in L. D. Landau and E. M. Lifshitz, {\it Theory of
Elasticity} 3rd Edition
\bibitem{kohn61}
W. Kohn, Phys. Rev. {\bf 123}, 1242 (1961)
\bibitem{bardeen61}
J. Bardeen, Phys. Rev. Lett. {\bf 6}, 57 (1961)
\bibitem{mahan}
See section 4.3.6 in G. D. Mahan, {\it Many-Particle Physics}, Third edition. Notice
formula (4.369) needs an additional term $\frac{2t}{i\omega}$ to be complete.
\bibitem{bonsall77}
L. Bonsall and A. A. Maradudin, Phys. Rev. B {\bf 15}, 1959 (1977);
H. Fukuyama, Solid State Commun. {\bf 17}, 1323 (1975)
\bibitem{cote}
R. C\^ot\'e {\it et al.} private communication.
\bibitem{He93}
Song He, P. M. Platzman, and B. I. Halperin, Phys. Rev. Lett. {\bf 71}, 777 (1993).
\end{thebibliography}
\end{document}